# CLUSTER VS SINGLE-SPIN ALGORITHMS – WHICH ARE MORE EFFICIENT?

N. Ito[†] and G.A. Kohring

*HLRZ at the KFA Jülich*
*Postfach 1913, D-5170 Jülich, Germany*
[†] *and Computing and Information Systems Center*
*Japan Atomic Energy Research Institute*
*Tokai, Ibaraki 319-11, Japan*

ABSTRACT

A comparison between single-cluster and single-spin algorithms is made for the Ising model in 2 and 3 dimensions. We compare the amount of computer time needed to achieve a given level of statistical accuracy, rather than the speed in terms of site updates per second or the dynamical critical exponents. Our main result is that the cluster algorithms become more efficient when the system size, $L^d$, exceeds, $L \sim 70$–$300$ for $d = 2$ and $L \sim 80$–$200$ for $d = 3$. The exact value of the crossover is dependent upon the computer being used. The lower end of the crossover range is typical of workstations while the higher end is typical of vector computers. Hence, even for workstations, the system sizes needed for efficient use of the cluster algorithm is relatively large.

## 1. Introduction

The phenomena of critical slowing down within computer simulations of statistical systems near a second-order phase transition has been known for many years. The basic problem stems from the difficulty encountered by the traditional Metropolis Monte Carlo method of flipping individual spins when the correlation length is very large. In such a case, any given spin will tend to align itself with those spins with which it is correlated, thus the problem of flipping a given spin becomes the problem of flipping the cluster of spins correlated with the given spin. This latter task can be quite time consuming when using a single-spin flip algorithm, because only small parts of a cluster are likely to be flipped during a single pass through the lattice. Obviously, it would be better to have an algorithm which can flip the entire cluster of correlated spins during a single pass through the lattice.

A suitable definition for a cluster of correlated spins was first given by Fortuin and Kasteleyn.[1] Coniglio and Klein later showed that there is a correspondence between the Fortuin and Kasteleyn clusters and Fisher's droplet ideas.[2] In 1983, Sweeny[3] first applied these ideas directly to the problem of critical slowing down in the Ising model. However, Sweeny's approach was difficult to realize in practice and was never widely used. In 1987, Swendsen and Wang [4] presented an alternative approach which was easier to implement and to generalize to systems other than the Ising model. In the Swendsen-Wang procedure, all possible Fortuin-Kastelyn clusters are identified and then each is flipped with probability 1/2. Following Swendsen and Wang several authors proposed that equivilent results could be achieved by simply identifying and flipping a single cluster chosen at random.[5] Since only that cluster which is to be flipped needs to be identified, this method should be faster and simpler than the Swendsen-Wang approach.

In the literature there have been many claims about the efficiency of the cluster methods vis-à-vis the traditional single-spin flip algorithms.[6] Most of these claims rest upon the calculation of the autocorrelation exponent $z$, which determines the asymptotic efficiency of the algorithm for very large systems. However, while the cluster methods were developing the traditional Metropolis Monte Carlo algorithms were also moving forward through the development of sophisticated vectorized algorithms.[7–9] Today, the typical time needed to update a single spin with a vector machine is on the order of a nanosecond. A more relevant question, then, is the amount of cpu time needed to obtain a given statistical accuracy for system sizes of practical importance. Here, we examine this question for both vector computers and scalar workstations. Our main result being, that even for scalar workstations, the system sizes where the cluster methods become more efficient are relatively large.

## 2. The algorithms

The spin models we wish to use for our comparisons are the simple Ising models in two and three dimension. These models consist of of $N$ spins, $S_i$, arranged on a simple cubic lattice ($L^d = N$) in $d$ dimensions and taking on the values, $\{\pm 1\}$. The spins interact only with their nearest neighbors. The traditional single-spin-flip Metropolis Monte Carlo method, proceeds as follows:

1) pick a spin, $S_i$, either randomly or systematically;
2) calculate the change in energy, $\Delta E$, which would occur if the spin were flipped to its opposite value;
3) if $\Delta E < 0$ then the spin flip is accepted otherwise the spin flip is accepted with probability $\exp(-\beta \cdot \Delta E)$, where $\beta$ is the inverse temperature of the system.

Since each of the $S_i$ take on only two values, memory space can be saved by packing more than one variable into a computer word, the so-called multi-spin coding technique.[7] The early practitioners of this technique worked with a single lattice, however it was latter realized that the updating speed could be greatly enhanced by simulating multiple lattices. [8,9] Basically, the idea is to simulate $B$ lattices simultaneously, on a computer with $B$ bits per word. Originally, the multiple lattice technique was limited to simulating $B$ different temperatures as well, however, this restriction has also been lifted. [8] For the 3-D Ising model, these techniques currently enable one to reach speeds of nearly 466 million site updates per second (466 Mups) on a single Cray-YMP processor, 1040 Mups on the JAERI and NEC's experimental MONTE-4 machine and 2190 Mups on Fujitsu VP2600/10. The MONTE-4 is designed and constructed based on the NEC SX3/41 with some additional vector pipelines, and other improvements including a 2.5 nsec clock cycle.[16] The reader interested in the details of this algorithm can find them in ref. 9, while an example implementation is given in Appendix B.

For the cluster-flip algorithms, one proceeds quite differently from the procedure given above. As stated in the introduction, one must identify a cluster of correlated spins and attempt to flip the entire cluster. The single cluster algorithm [5] works as follows:

1) pick a spin, $S_i$, at random as the first spin in the cluster;
2) if the neighboring spins are of the same sign as spin $S_i$, then add them to the cluster with probability $1 - e^{-2\beta}$, otherwise ignore them;
3) repeat 2) with the newly added spins until no new spins are added to the cluster;
4) flip all the spins in the cluster to their opposite value.

As mentioned in the introduction, this single cluster algorithm is easier to implement than the full cluster algorithm because there is no need to label all of the individual clusters. This procedure is more complicated than the single-spin algorithm, hence the updating speeds of are generally quite small, approximately 0.1 Mups.

Recently, it was shown, that the single cluster algorithm could be vectorized by vectorizing the cluster search over the active sites on the perimeter of the cluster.[10] The kernel of such a vectorized algorithm is given in Appendix C. Obviously, this algorithm will only be efficient if the vector length is "long enough". (Where the meaning of "long enough" is hardware dependent.) Fig. 1 shows a histogram of the vector length taken from a 2-D system with $L = 512$. The histogram is over 10,000 clusters taken at the critical temperature of the 2-D Ising model. In this example, the average vector length is 98, however, the most probable vector length is 19. This reflects the fact, that initially the cluster is quite small, hence there are very few points on the perimeter. As the cluster grows, so does the number of perimeter sites, however, at some point, the number of active sites on the perimeter reaches a maximum and then decreases until the cluster finally stops growing. Given, the wide distribution of vector lengths, the efficiency of this algorithm on a vector computer depends upon how well the compiler handles short vector loops and how well the hardware and software can handle random memory accesses. In the next section, we test the efficiency of this algorithm using vector and scalar computers.

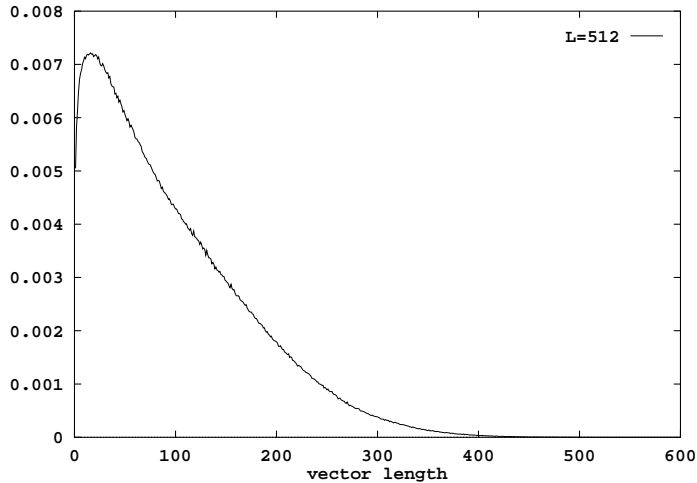

Fig. 1. Histogram of the vector length for a 2-D system with $L = 512$. The histogram was made at the critical point using 10,000 clusters.

## 3. Results

A comparison of the forementioned single-spin-flip and cluster-flip algorithms was made on vector and scalar computers. For the vector computers we chose the Cray-YMP/832 and the MONTE-4 two machines with very different hardware characteristics and for the scalar machines we chose the SUN Sparc workstation series.

When comparing the above algorithms, speed is not the only criterion, because the auto-correlation times are very different. The auto-correlation time, $\tau$, measures the statistical accuracy achievable with $n$ Monte Carlo sweeps through the lattice. Typically, the statistical error, $\epsilon$, is given by: $\epsilon \approx \sigma/\sqrt{(n/\tau)}$, where $\sigma$ is the standard deviation in some measured quantity. At the critical temperature, $T_c$, it is known that for the above algorithms, $\tau \sim L^z$. Where $z_{ss} \approx 2.0$, for the single-spin case,[11,12]. For the single-cluster algorithm, it is customary to multiply the correlation time in terms of cluster flips by the average relative cluster size. In this way, one obtains the correlation time per spin flip, which has a $z$ value of, $z_{sc} \approx 0.25$.[6] This practice is unfortunate, because one can only measure observables after each cluster has been flipped and not when a part of the cluster has been flipped, hence, to determine the required cpu resources, one must use the correlation time in terms of cluster flips. This correlation time is much larger than the one usually mentioned in the literature[6], because the average relative cluster size tends to decrease like $O(L^{2-d-\eta})$. Consequently, the effective $z$ value for the cluster updating is given by: $z_{eff} = z_{sc} + d + \eta - 2$. In 2-D, $\eta = 0.25$, which gives $z_{eff} = 0.50$ for the single-cluster algorithm.

Given the large speed differential and the large difference in $z$, the only reasonable criterion for comparing these two algorithms is by comparing the amount of cpu time needed to obtain a given level of statistical accuracy. Asymptotically, this cpu time, $t_{cpu}$, is $t_{cpu}^{ss} \propto L^{d+z_{ss}}$ and $t_{cpu}^{sc} \propto L^{d+z_{eff}}$. Hence, asymptotically, the cluster algorithm will be more efficient (i.e., require less cpu time) than the single-spin algorithm, however, the proportionality constants differ considerably, because the cluster algorithm is more complex than the single-spin algorithm, so that for small systems the single-spin algorithms will be more efficient. The question we want to ask here, is whether or not the system size at which the cross-over from single-spin efficiency to cluster efficiency is "reasonable".

When calculating $\tau$ there is one subtlety to be taken into account: different observables can have different values of $\tau$. In particular, even quantities like the energy and the susceptibility are expected to have $\tau$'s which are much different than odd quantities like the magnetization, although most people expect that $z$ should be independent of the observable in question. If the sampling interval of the quantity to be calculated, $n$ in terms of Monte Carlo steps, is much large than $\tau$, then the the auto-correlation time for all even quantities, should be the same. However, when using the integrated correlation time, $\tau_{int}$, as a measure of $\tau$, different observables can strongly couple to one or more fast modes, thus under estimating the true value of $\tau$. In the 2-D case, we find that estimates of $\tau_{int}$ for the energy converge to that of the magnetization squared, when measurements are done for $n \gg \tau_{int}$. For shorter runs, we observe a much smaller $\tau$ for the energy than for the magnetization squared, hence, our comparisons here are made with $\tau_{int}$ as estimated for the magnetization squared. In the 3-D case, we find a much smaller difference between the estimates of $\tau$ taken from the magnetization squared and the energy even for rather short runs.

In Fig. 2a is a plot of the cpu time necessary for $100 \cdot \tau_{int}$ measurements for the 2-D Ising model using single-spin and cluster algorithms on vector computers. (The raw data for all graphs is given in Appendix A.) For the single-spin algorithms we measure the speed in terms of Mups and then multiply this speed by the lattice size and 100 times the correlation time.[11,12] For the cluster algorithm, we measure the average time needed to perform 100 cluster updates, then multiply this by the correlation time in terms of Monte-Carlo sweeps. We used the energy auto-correlation time as measured with the technique recently introduced in ref. 12. In all cases, the energy auto-correlation time per spin and its corresponding critical exponent are in agreement with results previously reported.[6] The single-spin algorithm runs at a peak speed of 1290 Mups on the MONTE-4 and 720 Mups on the Cray-YMP (All times are for a single processor.). This speed difference reflects the difference in the clock cycle between the two machines. However, the peak speed of the cluster updating is approximately the same for each machine, 1.0 Mups. This reflects the fact that the MONTE-4 machine performs very poorly for short vector loops, the most probable situation for the cluster algorithm. For the smallest system sizes, the Cray-YMP is running at a speed of only 21 million operations per second, while for the largest system sizes this speed increases to over 150 million operations per second, or about 60% of the YMP's sustainable speed. Hence, the vectorized cluster algorithms are indeed running efficiently on vector machines.

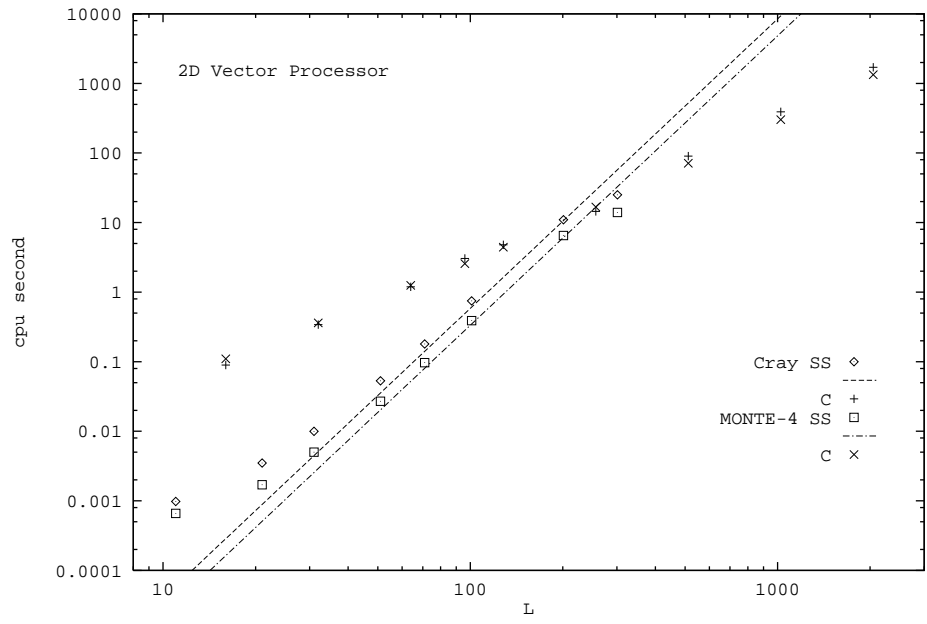

Fig. 2a. Cpu time need to make $100 \cdot \tau_{int}$ time steps with the 2-D Ising model using the Cray-YMP and the MONTE-4 vector computers.

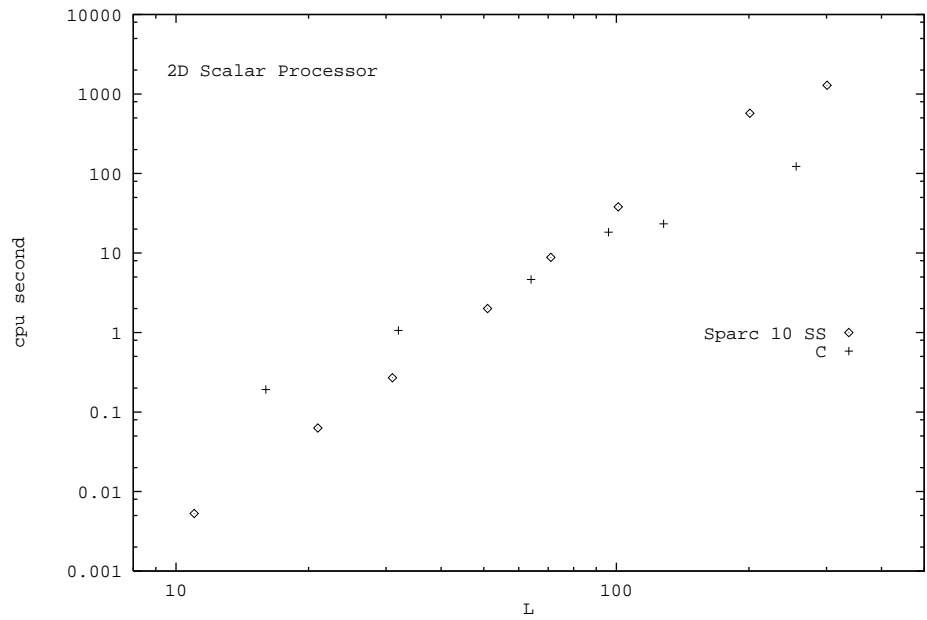

Fig. 2b. Same as (a) but for the SUN Sparc-10 scalar workstation.

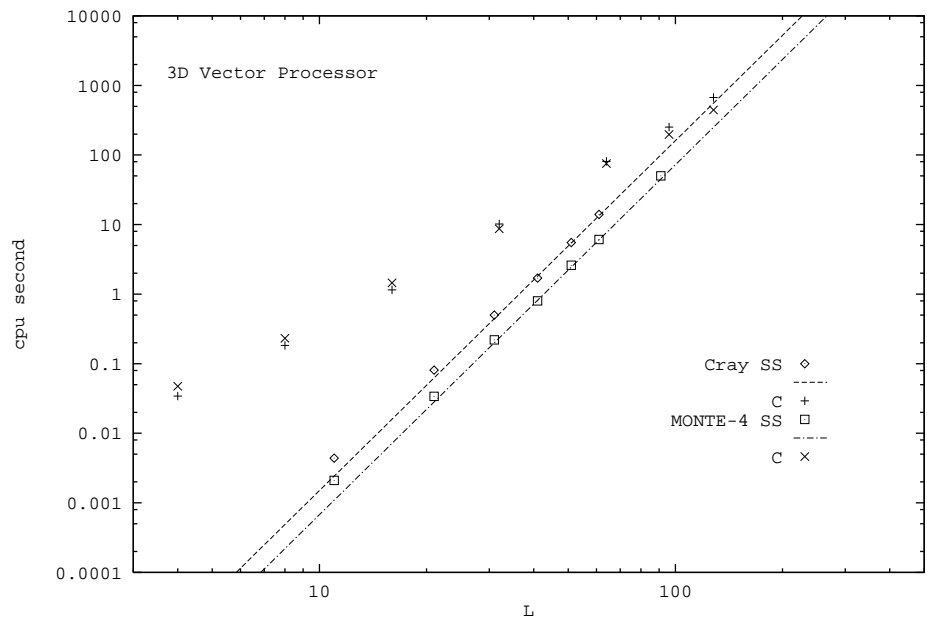

Fig. 3a. Cpu time need to make $100 \cdot \tau_{int}$ time steps with the 3-D Ising model using the Cray-YMP and the MONTE-4 vector computers.

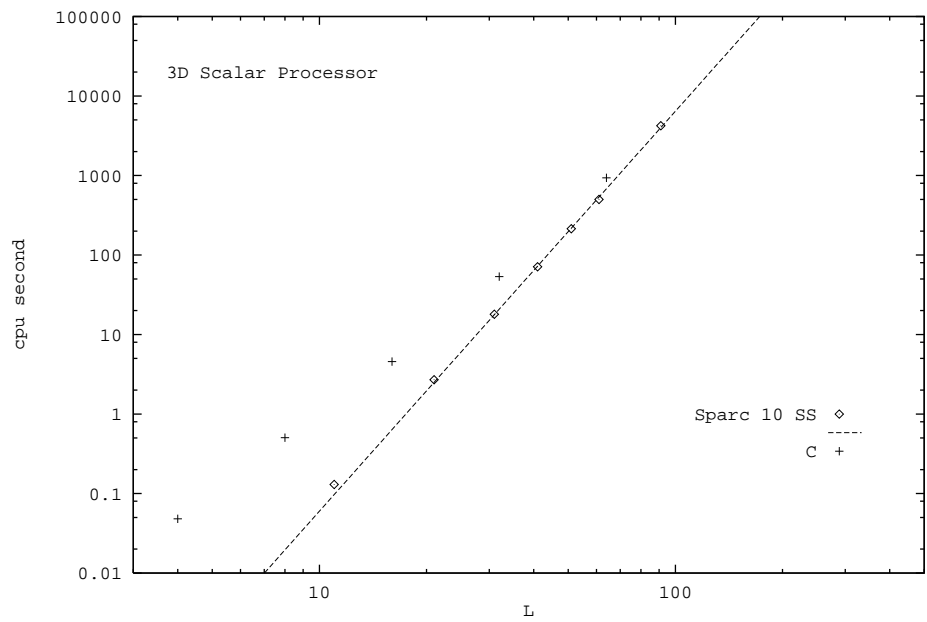

Fig. 3b. Same as (a) but for the SUN Sparc-10 scalar workstation.

From fig. 2a it can be seen that the cross-over size ranges from $L \approx 200$ for the Cray-YMP to $L \approx 300$ for the MONTE-4. Fig. 2b shows the results for the SUN-Sparc-10 workstation. For the SUN-Sparc-10 the cross-over size is $L \approx 70$. Hence, even on scalar workstations, the multi-spin coding algorithms are very efficient, and contrary to common belief, the cross-over sizes are nearly the same as for the vector computers.

These cross-over sizes are certainly small compared to the "world record" size of $L = 169,984$,[14] however, this world record size was only used to calculate the decay of the initial magnetization during the first 100 time steps and was not used for measuring equilibrium properties, which is the primary purpose of Monte Carlo methods. On the other hand, since most of the equilibrium properties of the 2-D Ising Model can be calculated analytically, it is more interesting to look at the behavior of the above algorithms for the 3-D situation, where the equilibrium properties cannot be calculated exactly .

Fig. 3a. shows the results for the 3-D Ising model on the vector computers. For the inverse critical temperature, $\beta_c$ we use the estimate: $\beta_c = 0.221656$. Here we see that the cross-over sizes are $L \approx 120$ for the Cray-YMP and $L \approx 200$ for the MONTE-4. Fig. 3b gives the results for the SUN-Sparc-10 workstations. Here the cross-over size is about $L \approx 80$. These sizes are somewhat smaller than for the 2-D Ising model, however, the "world record" size in 3-D is $L \approx 3072$, [14] hence the cross-over size for the 3-D systems are relatively larger than for the 2-D systems.

We can answer the question if these sizes are reasonable by looking at the high resolution studies of the 3-D Ising model performed by various groups in ref. 15. These papers have helped to set the current standard for accuracy in equilibrium Monte Carlo simulations. All three papers studied system sizes in the range $8 \leq L \leq 128$. The first two papers used single-spin flip algorithms, while the third paper used a single cluster algorithm. All three papers used roughly the same amount of computer resources and achieved nearly the same accuracy although they each analysed their data using different methods. Hence, these papers support the results found here, namely for systems up to $L \approx 100-200$, there is no clear advantage to be gained by cluster updating. Larger systems would not have helped these researchers much because they simply did not have enough computer time at their disposal to study larger systems. Thus, as far as calculating thermodynamic properties is concerned, the sizes at which the cluster algorithms are clearly more efficient than single-spin algorithms for the 3-D Ising model, are slightly larger than realistic, given the currently available computers and the currently known algorithms. This situation may change, given faster scalar processors or parallel computers. However, since most of the announced parallel computers include vector co-processors, the cross-over size may also increase with increasing speed of the vector co-processors.

Finally, it should be noted that single-spin algorithms have had a much longer time to develop than cluster-algorithms. Evertz's vectorized cluster algorithm is one step towards faster cluster algorithms and there is still hope for yet faster methods.

## Acknowledgements

We would like to thank H.G. Evertz and D. Stauffer for many help discussions related to this work.

# Appendix A

Table 1: Single-Spin-Flip Measurements of the 2-D Ising Model. The necessary CPU time for $100\tau_{int}$ Monte Carlo steps is given in seconds. The figures in "large $L$" row refer to the prefactor of $L^{4.16}$ (sec).

| size | $\tau_{int}$ | CRAY YMP | MONTE-4 | Sparc | S2 | S10 |
|---|---|---|---|---|---|---|
| $11 \times 12$ | 5.104 | 0.00098 | 0.00066 | 0.012 | 0.0077 | 0.0053 |
| $21 \times 22$ | 16.412 | 0.0035 | 0.0017 | 0.13 | 0.085 | 0.063 |
| $31 \times 32$ | 35.439 | 0.010 | 0.0050 | 0.62 | 0.39 | 0.27 |
| $51 \times 52$ | 97.49 | 0.053 | 0.027 | 4.6 | 2.9 | 2.0 |
| $71 \times 72$ | 203.8 | 0.18 | 0.097 | 19 | 12 | 3.8 |
| $101 \times 102$ | 422.9 | 0.75 | 0.39 | 79 | 50 | 38 |
| $201 \times 202$ | 1931 | 11 | 6.5 | 1455 | 906 | 572 |
| $301 \times 302$ | 4370 | 25 | 14 | 3304 | 2061 | 1286 |
| large L | $0.020L^{2.16}$ | $2.8 \times 10^{-9}$ | $1.6 \times 10^{-9}$ | $3.6 \times 10^{-7}$ | $2.3 \times 10^{-7}$ | $1.4 \times 10^{-7}$ |

Table 2: Single-Spin-Flip Measurements of the 3-D Ising Model. The necessary CPU time for $100\tau_{int}$ Monte Carlo steps is given in seconds. The figures in "large $L$" row refer to the prefactor of $L^{5.03}$ (sec).

| size | $\tau_{int}$ | CRAY YMP | MONTE-4 | Sparc | S2 | S10 |
|---|---|---|---|---|---|---|
| $11 \times 11 \times 12$ | 9.461 | 0.0044 | 0.0021 | 0.33 | 0.20 | 0.17 |
| $21 \times 21 \times 22$ | 32.979 | 0.081 | 0.034 | 7.9 | 4.9 | 4.0 |
| $31 \times 31 \times 32$ | 72.13 | 0.50 | 0.22 | 56 | 34 | 28 |
| $41 \times 41 \times 42$ | 122.3 | 1.7 | 0.80 | 201 | 124 | 101 |
| $51 \times 51 \times 52$ | 191.2 | 5.5 | 2.6 | 660 | 408 | 332 |
| $61 \times 61 \times 62$ | 274.0 | 14 | 6.1 | 1625 | 1015 | 811 |
| $91 \times 91 \times 92$ | 683 | 100 | 50 | 14025 | 8447 | 6749 |
| large L | $0.065L^{2.03}$ | $1.4 \times 10^{-8}$ | $6.3 \times 10^{-9}$ | $1.7 \times 10^{-6}$ | $1.0 \times 10^{-6}$ | $8.2 \times 10^{-7}$ |

Table 3: Single-Cluster-Flip Measurements of the 2-D Ising Model. The necessary CPU time for $100\tau_E$ Monte Carlo steps is given in seconds.

| size | $\tau_E$ | $<c>/L^d$ | CRAY YMP | MONTE-4 | Sparc | S2 | S10 |
|---|---|---|---|---|---|---|---|
| $16 \times 16$ | 2.4 | 0.565 | 0.0895 | 0.110 | 0.864 | 0.384 | 0.192 |
| $32 \times 32$ | 3.8 | 0.499 | 0.341 | 0.361 | 4.71 | 2.09 | 1.064 |
| $64 \times 64$ | 5.4 | 0.406 | 1.19 | 1.25 | 21.1 | 9.45 | 4.64 |
| $96 \times 96$ | 6.5 | 0.386 | 3.05 | 2.57 | 68.9 | 33.9 | 18.3 |
| $128 \times 128$ | 7.6 | 0.346 | 4.78 | 4.42 | 114 | 56.1 | 23.3 |
| $256 \times 256$ | 10.5 | 0.275 | 14.5 | 16.7 | 531 | 280 | 123 |
| $512 \times 512$ | | 0.252 | 90.1 | 71.3 | | | |
| $1024 \times 1024$ | | 0.208 | 390 | 301 | | | |
| $2048 \times 2048$ | | 0.173 | 1705 | 1335 | | | |
| large L | $0.70L^{0.49}$ | $1.20L^{-0.25}$ | | | | | |

Table 4: Single-Cluster-Flip Measurements of the 3-D Ising Model. The necessary CPU time for $100\tau_E$ Monte Carlo steps is given in seconds.

| size | $\tau_E$ | $<c>/L^d$ | CRAY YMP | MONTE-4 | Sparc | S2 | S10 |
|---|---|---|---|---|---|---|---|
| $4 \times 4 \times 4$ | 2.4 | 0.383 | 0.0343 | 0.0473 | 0.264 | 0.12 | 0.048 |
| $8 \times 8 \times 8$ | 5.6 | 0.195 | 0.183 | 0.232 | 2.46 | 1.01 | 0.504 |
| $16 \times 16 \times 16$ | 12.7 | 0.0967 | 1.15 | 1.45 | 21.3 | 9.14 | 4.57 |
| $32 \times 32 \times 32$ | 34.1 | 0.0473 | 10.2 | 8.7 | 237 | 141 | 53.5 |
| $64 \times 64 \times 64$ | 86.6 | 0.0232 | 80.9 | 73.3 | 3204 | 1931 | 935 |
| $96 \times 96 \times 96$ | | | 252 | 197 | | | |
| $128 \times 128 \times 128$ | | | 669 | 444 | | | |
| large L | $0.38L^{1.30}$ | $1.68L^{-1.03}$ | | | | | |

## Appendix B

The following subroutine updates a 2-D lattice using the single-spin, multi-spin coding technique of ref. 9. This subroutine is optimized for the Cray-YMP.

```
      SUBROUTINE SU2DSK(ISTEP,L1,L2,IS,IRD,IRLST,IX1,IX2)
      DIMENSION IS((-L1+1):(L1*(L2+1)))
      DIMENSION IRD(L1*L2)
      DIMENSION IX1(0:IRLST),IX2(0:IRLST)
      NSYS=L1*L2
      LS=L1
      DO 10 IMCS=1,ISTEP
      CALL RNDO2I(NSYS,IRD)
      IFIRST=1
CDIR$ IVDEP
   40 DO 20 I=-ls+1,0
   20 IS(I)=IS(I+nsys)
CDIR$ IVDEP
      DO 30 I=NSYS+1,NSYS+LS
   30 IS(I)=IS(I-nsys)
CDIR$ IVDEP
      DO 50 IJ=IFIRST,NSYS,2
         IST=IS(IJ)
         I1=IS(IJ+1)
         I2=IS(IJ-1)
         I3=IS(IJ+L1)
         K1=IEOR(I1,I2)
         K2=IAND(I1,I2)
         J2=IEOR(K1,I3)
         K3=IAND(K1,I3)
         J1=IOR(K2,K3)
         J1=ieor(ist,j1)
         J2=ieor(ist,j2)
```

```
      I4=IS(IJ-L1)
      I4=IEOR(I4,IST)
      IRT=IRD(IJ)
      IX1T=IX1(IRT)
      IX2T=IX2(IRT)
      K2=IEOR(I4,IX2T)
      K1T=IAND(I4,IX2T)
      K1=IOR(K1T,IX1T)
      ID=IOR(J1,K1)
      K4=IAND(J2,K2)
      ID=IOR(ID,K4)
      IS(IJ)=IEOR(IS(IJ),ID)
50    CONTINUE
      IF(IFIRST.EQ.1)THEN
        IFIRST=2
        GOTO 40
      END IF
10    CONTINUE
      RETURN
      END
```

## Appendix C

The following subroutine updates a 2-D lattice using the cluster method. It is far from optimal in terms of storage, but it is well optimized for speed.

```
      subroutine evertz(prob1,prob2)
C
      implicit none
      integer L,V
      parameter (L=128,V=L*L)
      integer list(0:V),cluster(0:V),i
      real    lat(0:V),prob1,prob2,site_0
      integer last,first,listend,site_i,n_site,,cnum
      data cnum/0/
      save cnum
      common /LATTICE/lat,list,cluster
C
      cnum = cnum + 1
      listend=1
      last = 0
      list(1)= ifix(float(V)*ranf())
      site_0 = lat(list(1))*prob2
      do while (listend .gt. last)
         first = last + 1
         last = listend
CDIR$ IVDEP
         do 10 i=first,last
            site_i = list(i)
            n_site = mod(site_i+1,V)
            if ( cluster(n_site) .ne. cnum) then
               if (ranf() .lt. (prob1 + site_0*lat(n_site) ) ) then
                  cluster(n_site) = cnum
                  listend=listend+1
                  list(listend)=n_site
```

```
                  lat(n_site) = -lat(n_site)
               end if
            end if
 10      continue
CDIR$ IVDEP
         do 20 i=first,last
            site_i = list(i)
            n_site = mod(V-1+site_i,V)
            if ( cluster(n_site) .ne. cnum ) then
               if (ranf() .lt. (prob1 + site_0*lat(n_site) ) ) then
                  cluster(n_site) = cnum
                  listend=listend+1
                  list(listend)=n_site
                  lat(n_site) = -lat(n_site)
               end if
            end if
 20      continue
CDIR$ IVDEP
         do 30 i=first,last
            site_i = list(i)
            n_site = mod(site_i+L,V)
            if ( cluster(n_site) .ne. cnum ) then
               if (ranf() .lt. (prob1 + site_0*lat(n_site) ) ) then
                  cluster(n_site) = cnum
                  listend=listend+1
                  list(listend)=n_site
                  lat(n_site) = -lat(n_site)
               end if
            end if
 30      continue
CDIR$ IVDEP
         do 40 i=first,last
            site_i = list(i)
            n_site = mod(V-L+site_i,V)
            if ( cluster(n_site) .ne. cnum) then
               if (ranf() .lt. (prob1 + site_0*lat(n_site) ) ) then
                  cluster(n_site) = cnum
                  listend=listend+1
                  list(listend)=n_site
                  lat(n_site) = -lat(n_site)
               end if
            end if
 40      continue
C
      end do
C
      return
      end
```